\documentclass[10pt, conference, letterpaper]{IEEEtran}

\usepackage{caption}

\usepackage{subcaption}
\usepackage{amsmath,mathtools}
\usepackage{amsfonts}
\usepackage{amssymb}
\usepackage{amsthm}
\usepackage{algorithm}
\usepackage{algpseudocode}
\usepackage{booktabs} 
\usepackage{fancyhdr}
\usepackage{xcolor}
\usepackage{multicol}
\usepackage{graphicx}
\usepackage{caption}
\usepackage{subcaption}  
\IEEEoverridecommandlockouts
\usepackage{soul}
\usepackage{pifont}
\usepackage[breaklinks=true]{hyperref}
\usepackage{boxedminipage}
\usepackage{tikz}
\usepackage{multirow}
\usepackage{array}
\usepackage{enumitem,kantlipsum}
\usepackage{comment}
\usetikzlibrary{patterns}
\usepackage{tikzscale}

\usepackage{url}

\newcommand{\circled}[1]{\text{\textcircled{\scriptsize #1}}}

\begin{document}
\vspace{-0.2in}
\title{\LARGE{O-DSS: An Open Dynamic Spectrum Sharing Framework for Cellular-Radar Coexistence in Mid-band Frequencies}
\thanks{Authors would like to acknowledge the funding support from the National Science Foundation (NSF) awards \#2443035, \#2120411, \#2526490 and \#2526493.}
}
\author{\IEEEauthorblockN{Azuka Chiejina\IEEEauthorrefmark{1}, 
Divyadharshini Muruganandham \IEEEauthorrefmark{2}, Vini Chaudhary \IEEEauthorrefmark{3}, Kaushik Chowdhury\IEEEauthorrefmark{2}  and Vijay K. Shah\IEEEauthorrefmark{1}}
\IEEEauthorblockA{\IEEEauthorrefmark{1}\textit{NextG Wireless Lab}, NC State University, USA, \IEEEauthorrefmark{2} The University of Texas at Austin, USA}
\IEEEauthorrefmark{3} Mississippi State University, USA
\IEEEauthorblockA{Email: \{ajchieji, vijay.shah\}@ncsu.edu, \{muruganandham.d, Kaushik\}@utexas.edu, vchaudhary@cse.msstate.edu}
}

\newcommand{\Vijay}[1]{\noindent\textcolor{red}{#1}}

\maketitle
\begin{abstract}
The growing demand for mid-band spectrum necessitates efficient Dynamic Spectrum Sharing (DSS) to ensure coexistence between cellular networks and incumbent radar systems. Existing Spectrum Access System (SAS) frameworks rely on fixed Environmental Sensing Capability (ESC) sensors, which are latency-prone and inflexible. This paper introduces \textbf{O-DSS}, an O-RAN-compliant, Machine Learning (ML)-driven DSS framework that enables real-time cellular-radar coexistence in mid-band frequencies 
with shipborne and fast-moving airborne radars.
O-DSS integrates radar detection from low-overhead Key Performance Metrics (KPMs) with spectrogram-based localization to drive fine-grained RAN control, including PRB blanking and radar-aware MCS adaptation. Deployed as a modular xApp, O-DSS achieves \(\sim60\)~ms detection and \(\sim700\)~ms evacuation latencies, outperforming existing baselines. Evaluations across simulations and Over The Air (OTA) testbed show that O-DSS ensures robust incumbent protection while maintaining cellular performance by achieving radar detection of \(\geq 99\%\) at SINR \(\geq -4\)~dB and localization recall of \(\geq 95\%\) at SINR \(\geq 8\)~dB. 
\end{abstract}


\vspace{-0.02in}
\section{Introduction}
\vspace{-0.02in}
The US National Spectrum Strategy (NSS)~\cite{NSS} has identified the Radio Frequency (RF) spectrum as one of ``\textit{a nation’s most important national resources}''. Federal governments, consumers, and businesses at every level rely on the RF spectrum to complete a significant number of tasks from mundane to the critical. As a result of ongoing innovations in wireless technologies, the demand for spectrum access is growing rapidly \cite{1270548}. Sub-6GHz spectrum is now congested and ``greenfield'' spectrum is harder to find. Furthermore, the traditional approach of re-purposing existing federal wireless systems and fixed satellite service systems is becoming challenging \cite{Intro_4_1}. To address the growing spectrum demand within the US, the NSS identifies five spectrum bands totaling $2,786$ megahertz of spectrum for in-depth study to determine the suitability of potential repurposing to address the nation's ever-evolving needs. These spectrum bands are a mix of federal and shared Federal/Non-Federal bands, with an emphasis on \textbf{mid-band frequencies}, such as, lower $3$ GHz ($3.1 - 3.45$ GHz), $5030 - 5091$ GHz, and $7.125 - 8.4$ GHz, which support numerous federally operated fixed/mobile air, shipborne or terrestrial radar systems for various operations.

Dynamic spectrum sharing (DSS) is imperative to meet these growing demands for spectrum access in such mid-band frequencies~\cite{Intro_4,Intro_5}. The Citizen Broadband Radio Service (CBRS) band ($3.55-3.7$ GHz) is an example band that has been opened up in the US for sharing between federal incumbent systems and Priority Access License (PAL)/General Authorized Access (GAA) users. In order to protect incumbent users, different agencies such as the Federal Communications Commission (FCC), Wireless Innovation Forum and National Telecommunications and Information Administration (NTIA) have created a series of rules and policies controlling hierarchical access in the CBRS band \cite{fcc2015rno, fcc2024nprm,sanders2017procedures} by using a configuration of deployed sensors called Environmental Sensing Capability (ESC) to detect incumbent in the CBRS band and report to a commercially maintained Spectrum Access System (SAS) database. Thereafter, the CBRS band devices that seek to use unused channels communicate with the SAS for subsequent resource allocation \cite{SASdemo_1,SASdemo_2}. 

\begin{figure}[]
    \centering
\includegraphics[width=0.48\textwidth]{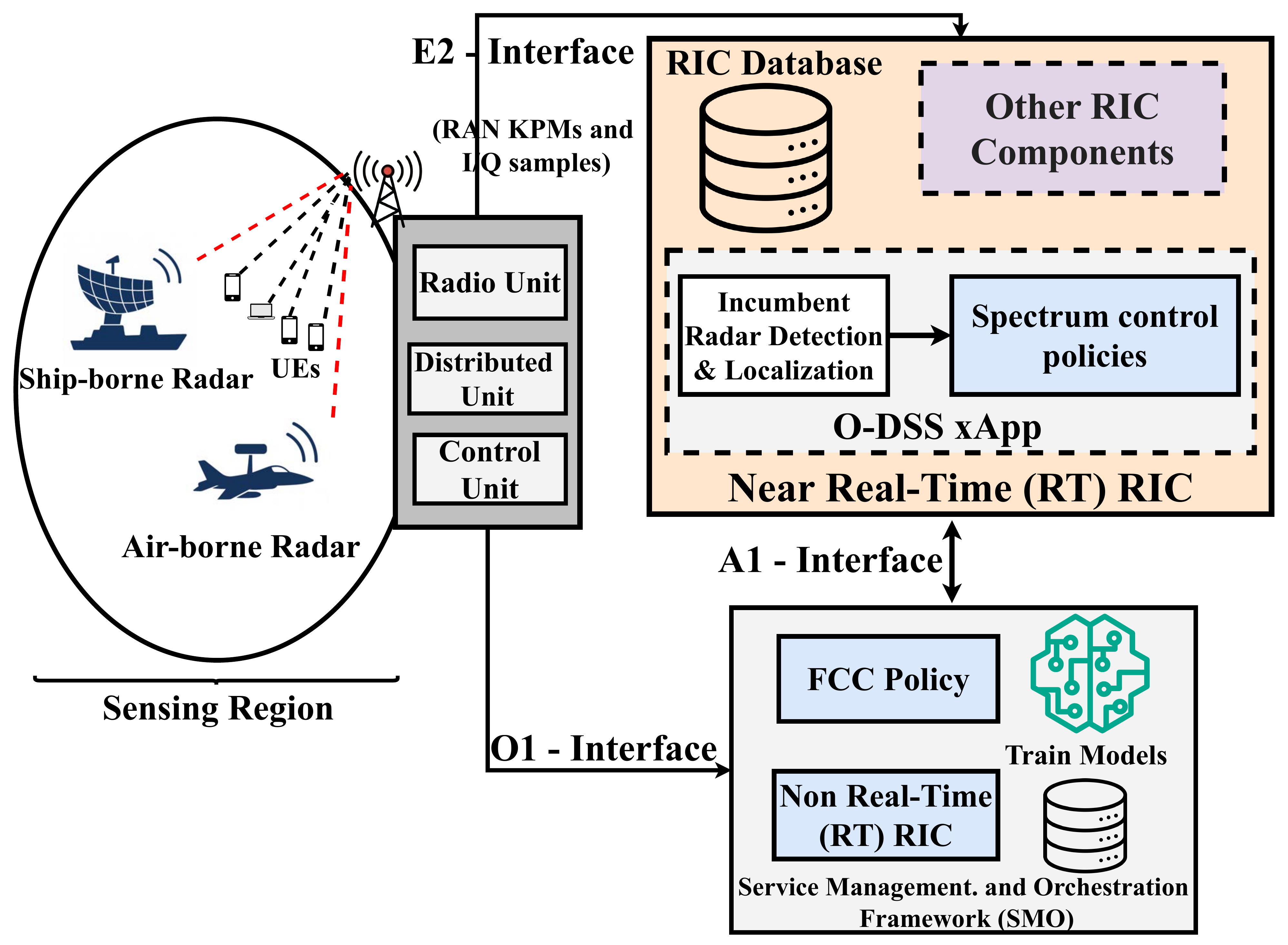}
 \caption{Overview of O-DSS Framework.}
    \label{fig:odss_overview}
    \vspace{-0.2in}
\end{figure}
\noindent
While SAS-based DSS solution works reasonably well for fixed terrestrial or slowly moving federal systems (shipborne radar), it has a number of critical limitations that makes it inadequate, inefficient, or simply infeasible to enable sharing in complex mid-band frequencies \cite{SAS_limitation}. 
We identify the following three key limitations.
    
 \noindent $\bullet$  \textit{\textbf{Reliance on external and dedicated ESC sensor network.}} SAS requires a strategically deployed network of external ESC sensors to collect In-phase (I) and Quadrature (Q) (I/Q) samples in the shared spectrum for spectrum sensing purposes. ESC sensors are (i) \textit{expensive} requiring federal certifications, are developed and deployed by private companies, (ii) \textit{sparsely deployed}, which is inadequate for federal systems with greater flexibility of motion, and (iii) \textit{susceptible to interference}, so cellular towers must be hundreds of kilometers away from the coast.

 \noindent $\bullet$ \textit{\textbf{Extremely slow system.}} SAS, with the aid of the ESC sensors was designed to achieve operational latency of $60$ seconds for incumbent detection and $240$ seconds for channel evacuation time \cite{sahoo2019study}, respectively. Although this might be suitable for shipborne radars, this latency is inadequate for fast, mobile airborne radars. Even with recent advances, as shown in \cite{reus2023senseoran}, response times still remain over 800 ms.

\noindent $\bullet$ \textit{\textbf{Absence of intelligence and mechanisms for sharing.}} Existing SAS-like solutions largely lack intelligent spectrum sharing capabilities and employ the worst-case assumptions for incumbent protection and interference management. Besides, current CBRS regulations lack specific policies, mechanisms or guidance on how the spectrum should be shared. Such factors severely limit the usage of the shared spectrum  for the co-existing  secondary services, including cellular networks.

\noindent \textbf{Proposed Solution: O-DSS Framework.} We propose a novel \textit{Open Dynamic Spectrum Sharing (O-DSS) framework} designed to address the limitations of traditional SAS-like DSS solutions and enable a fast, intelligent and reliable spectrum coexistence between cellular networks and Federal/Non-Federal incumbents (e.g., radar systems) in mid-band frequencies. The O-DSS framework leverages the standard-compliant, virtualized, disaggregated and programmable O-RAN architecture, which supports near-real-time RAN monitoring and control via Artificial Intelligence and Machine Learning (AI)/(ML) driven closed-loop decision making (See Section \ref{odss}). As shown in Figure~\ref{fig:odss_overview},  O-DSS has been designed to be deployed as an xApp --- referred to as the O-DSS xApp, that integrates ML techniques to detect and further localize the spectrum regions utilized by the incumbent radar transmission. It then employs  rule-based spectrum control mechanisms (such as, Physical Resource Block (PRB) blanking and Modulation and Coding Scheme (MCS) adaptation) to dynamically reconfigure the RAN, enabling harmonious cellular-radar coexistence in the shared band.  As a case study, we focus on the CBRS  band ($3.55 - 3.7$ GHz) with shipborne (Section V) and airborne radars (Section VI).
In summary, this paper makes the following key technical contributions:

  $\bullet$ We propose \textbf{O-DSS} --- an O-RAN based dynamic spectrum sharing framework for enabling cellular–radar coexistence in mid-band frequencies, with a case study of the CBRS and lower 3 GHz bands. O-DSS enables cellular base stations to act as pseudo-sensors, and eliminates the need for dedicated external ESC sensors, which has been a huge bottleneck/overhead in existing SAS-like DSS systems.
  
  $\bullet$ At the heart of the O-DSS framework is the \textbf{O-DSS xApp}, which consists of two major modules: (i) the \textit{Incumbent radar detection and spectrum localization module}, that operates in two distinct modes ---  (i.a) Mode 1: \textit{DNN-based incumbent radar detection} using RAN Key Performance Metrics (KPMs), (i.b) Mode 2: \textit{Yolo-based radar localization \footnote[2]{We do not localize the radar transmitter, rather, we localize the spectrum region being utilized (or affected) by the incumbent radar transmission.}} using spectrograms,
  (ii) the \textit{Dynamic spectrum control module}, which forms a closed-loop control layer that can leverage insights from the detection and localization modules to execute real-time RAN reconfiguration mechanisms like PRB blanking, radar-aware MCS optimization, and frequency switching.
    
 $\bullet$ We prototype the O-DSS framework through simulations and testbed using an open-source cellular stack \texttt{srsRAN} with ZeroMQ (ZMQ)-based RF simulation and Commercial Off-The-Shelf (COTS) SDRs. We perform experiments to understand how to protect incumbents, evaluate the impact of radar signals on cellular network performance, and show results that validate the feasibility of sharing spectrum between telecom and radar signals. 
 Our experimental evaluation demonstrates that O-DSS achieves \textbf{\(\geq 99\%\)} detection accuracy within \textbf{$\sim$60 ms} at Signal-to-Interference-plus-Noise Ratio (SINR) $\geq -4\,$dB, and \textbf{$\geq 95\%$} localization accuracy within \textbf{$\sim$700 ms} at SINR $\geq 8\,$dB. By relying on KPMs to detect external interference, we avoid continuous in-phase and phase-quadrature (I/Q) sample acquisition, thereby reducing E2 overhead and overall latency.

$\bullet$ 
To the best of our knowledge, this is the first initial study to consider air-borne radars in the context of cellular–radar spectrum coexistence.
We develop comprehensive datasets comprising both ship-borne and air-borne radars for training the ML models used in the proposed O-DSS framework: (i) RAN KPMs from an open-source cellular RAN stack sharing spectrum with a system transmitting radar, and (ii) I/Q samples and spectrograms capturing cellular–radar coexistence with the Type 2 P0N \#2 radar signal
under various conditions and capturing both ship-borne and air-borne radar characteristics. This radar type, known for its side lobe emissions \cite{sanders2013radar} and severe effect on cellular network performance, allows us to rigorously assess coexistence performance. 


\vspace{-0.02in}
\section{Related Works}
\vspace{-0.02in}
Several works in literature have explored spectrum sharing between cellular and radar systems, particularly for CBRS band. Most existing efforts have focused on leveraging ML techniques for incumbent radar detection using I/Q samples and spectrogram representations. In \cite{soltani2022finding}, the authors propose ESC+, an enhanced ESC sensor that detects both radar and cellular signals using spectrograms processed by a modified YOLOv3 neural network. Their methodology employs a coarse-to-fine detection strategy tailored to the short-duration nature of radar pulses. \cite{lees2019deep} demonstrates that deep learning methods outperform classical techniques in spectrogram-based signal detection. In \cite{caromi2019detection,caromi2021deep}, the authors utilize SVMs and CNNs for radar signal detection, applying both raw magnitudes and spectrogram inputs. \cite{sarkar2021deepradar} introduces a real-time YOLO-based ESC sensor that achieves 99\% detection accuracy and estimates radar bandwidth under noisy conditions.

Majority of efforts in literature emphasize radar signal detection without extending into the design of actionable spectrum coexistence mechanisms. Some studies have addressed spectrum sharing strategies within the CBRS context. \cite{tarver2019enabling} proposes Listen-Before-Talk (LBT) mechanisms, which may require significant modifications to 3GPP standards. In \cite{krishnan2017coexistence}, a power control algorithm is developed to facilitate radar-region communication by minimizing cellular interference. 
While \cite{lacava2025dapps} uses PRB blanking for sharing the spectrum in the presence of a uniform interferer requiring accurate noise floor estimation, \cite{ghosh2024sparc} uses similar PRB blanking technique for multi-site spectrum management. \cite{baldesi2022charm} presents various base station response strategies for LTE-Wi-Fi coexistence, opting for non-coexistence mode through frequency switching, while noting that coexistence  mode (i.e., intra-channel coexistence) approaches were beyond the scope of their work. 

Finally, compared to legacy, proprietary, black-box RAN, the O-RAN architecture introduces programmability and modularity, enabling more sophisticated coexistence mechanisms. Several works \cite{smith2021ran,reus2023senseoran,damnjanovic2024spectrum,acharya2023mitra,baldesi2022charm,gopal2024prosas,smith2024enabling} propose utilizing O-RAN to support spectrum sharing across various RF environments, offering architectural frameworks and some proof-of-concept demonstrations. Among all these, the most closely related work to ours is \cite{reus2023senseoran}, which also employs base stations as sensing entities within the proposed O-RAN based SenseORAN framework for radar detection. However, their work is limited to detecting land-based or shipborne radar and does not account for the unique challenges posed by dynamic airborne radar systems. These systems often exhibit higher mobility, varied altitudes, and distinct Doppler profiles, necessitating lower-latency detection and fine-grained control mechanisms.

\vspace{-0.02in}
\section{O-DSS for Cellular-Radar Coexistence}
\label{odss}
\vspace{-0.02in}
\subsection{Background on O-RAN}
\vspace{-0.02in}
The O-RAN architecture is revolutionizing cellular network design through the disaggregation and softwarization of the traditional 5G RAN stack. In this architecture, the Central Unit (CU) hosts the upper layers of the protocol stack, the Distributed Unit (DU) handles portions of the higher layer processing including high PHY layer components, and the Radio Unit (RU) is responsible for lower PHY layer operations. A fundamental feature of O-RAN is its support for ML-driven, closed-loop control enabled via RAN Intelligent Controllers (RICs)~\cite{tripathi2025fundamentals}. The Near-RT RIC can perform dynamic spectrum decision making on the order of sub-seconds. 

O-RAN’s standardized interfaces and real-time telemetry access make it ideal for spectrum-aware control. Unlike static SAS-based systems that suffer from coarse updates and lack of RAN flexibility, O-RAN empowers fine-grained, low-latency decisions at the network edge. These capabilities underpin our proposed O-DSS framework, enabling reliable and responsive coexistence with federal radar systems in mid-band spectrum.

\begin{figure}[!ht]
    \centering
    \includegraphics[width=0.48\textwidth]{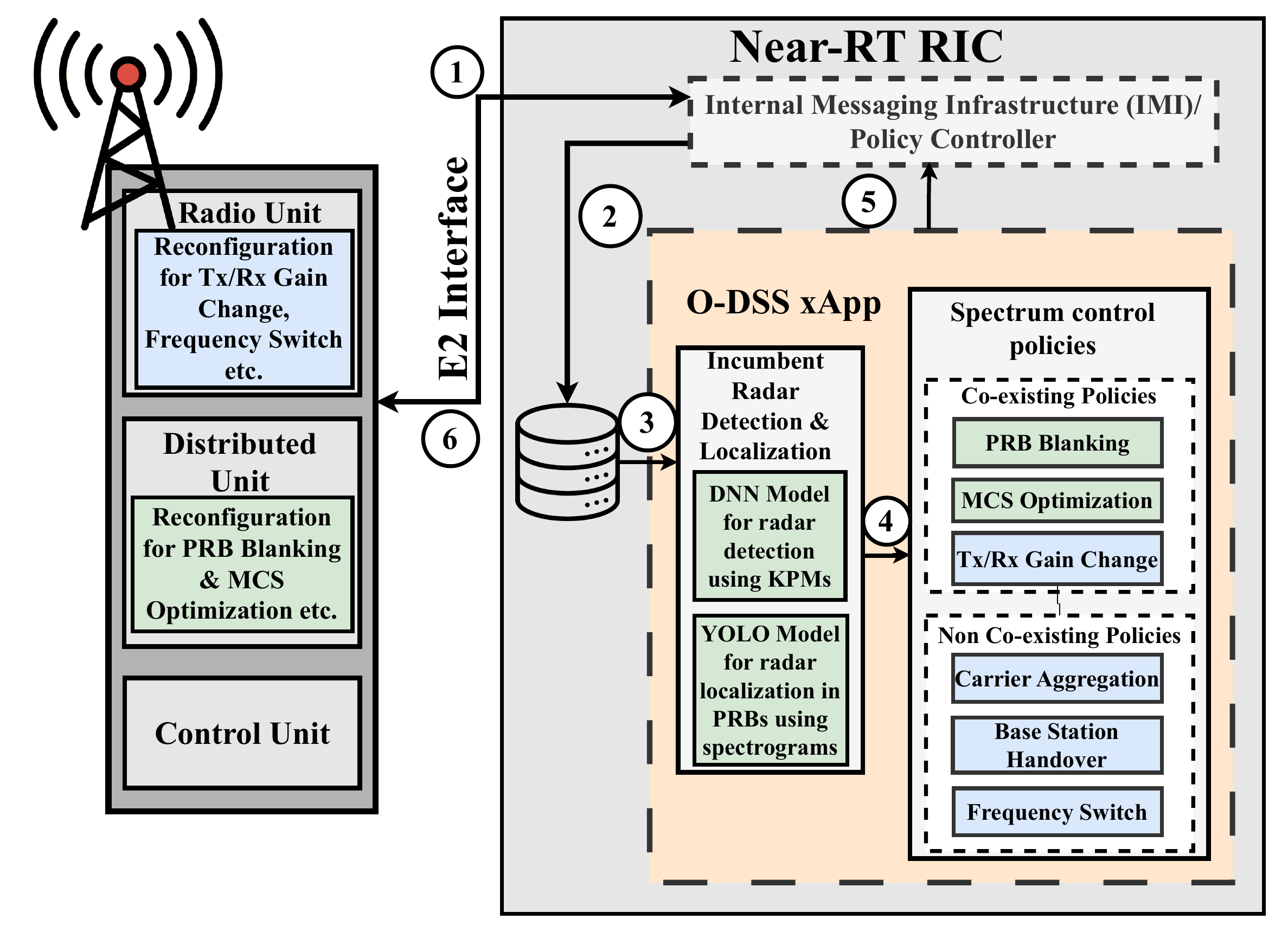}
    \caption{O-DSS system framework. For the purpose of this paper, we focus on performing spectrum coexistence as highlighted using sub-blocks in green while we would discuss how our system can be extended for the non-coexisting policies.}
    \label{fig:odss}
    \vspace{-0.2in}
\end{figure}

\vspace{-0.02in}
\subsection{Overview of the O-DSS Framework}
\label{odss_overview}
\vspace{-0.02in}
Figure~\ref{fig:odss} presents the O-DSS framework, which leverages the O-RAN architecture to enable low-latency, accurate incumbent detection and flexible spectrum control for both coexistence and non-coexistence scenarios in mid-band frequencies.


\textbf{O-DSS xApp Walk-through:} The operation of the O-DSS xApp can be summarized as follows:

\noindent\textbf{\circled{1}} The E2 interface acts as the communication bridge between the RAN and the Near-RT RIC. It connects to the policy controller, which is responsible for retrieving all telemetry data, performing preprocessing, and storing it in a centralized database. The telemetry data requested by the O-DSS xApp includes KPMs and/or I/Q samples, with configurable periodicity and data granularity.

\noindent\textbf{\circled{2}} The collected telemetry data is stored in a database, enabling persistent access by the xApp and other potential applications. This shared storage architecture ensures modularity and reusability of data across multiple xApps operating within the RIC environment.

\noindent\textbf{\circled{3}} The data is fed into the \textit{Incumbent radar detection and localization module} of the xApp, which operates in two modes depending on the type of input data (KPMs or I/Q samples). The module applies ML-based inference to determine the presence and characteristics of radar activity. These insights directly influence subsequent spectrum control decisions.

\noindent\textbf{\circled{4}} Based on the output from the detection/localization module, the xApp dynamically selects and triggers a spectrum control policy. These policies are categorized into:
\vspace{-0.05in}
    \begin{itemize}
        \item \textbf{Coexistence policies:} Techniques designed to allow simultaneous operation of cellular and radar systems, such as PRB blanking/unblanking and MCS optimization.
        \item \textbf{Non-coexistence policies:} Spectrum avoidance techniques such as frequency switching, or Base station turn off, similar to SAS-like proposed approaches and in~\cite{baldesi2022charm}.
    \end{itemize}
\noindent\textbf{\circled{5} \& \circled{6}} The final set of control commands generated by the xApp includes both the selected spectrum policy and any updates to the telemetry data request (e.g., to adjust type or periodicity). These commands are passed back to the policy controller, which relays them to the RAN through the E2 interface. On the RAN side, we have extended the \texttt{srsRAN} software stack to enable real-time reconfiguration in accordance with the selected policies, while maintaining uninterrupted cellular communication.

\vspace{-0.05in}
\subsection{Incumbent Radar Detection \& Localization Module} \label{rad_detection}
\vspace{-0.05in}
The \textit{Incumbent Radar Detection \& Localization Module} within the O-DSS xApp operates in two distinct modes, depending on data type requested by the detection pipeline.

\noindent
\textbf{$\blacksquare$ Mode 1: Low Data Overhead Radar Detection Using RAN KPMs:}
This mode serves as the default operating state of O-DSS system. During each observation period of duration $t_s$, the xApp collects $M$ KPMs from the RAN. To increase robustness and temporal awareness, the system can optionally stack KPMs from $N$ consecutive time windows, forming a total observation window of $t_o = N \times t_s$, with a combined input of $K = N \times M$ KPMs.

This stacked KPM data is fed into a DNN model trained offline on datasets collected under both radar-present and radar-absent conditions, across varying SINR levels. DNNs have been widely adopted for wireless network inference tasks using KPMs, demonstrating strong classification performance in prior work~\cite{10622798, 10556189, chiejina2024system}. Our choice to adopt this method is therefore well motivated. Among the KPMs that can be selected for training the model based on domain knowledge include the SINR, Throughput, Block Error Rate (BLER), MCS, and Buffer Status Report (BSR).

If the model predicts the absence of radar, the system remains in Mode 1, continuing to monitor the RAN via KPMs. If radar is detected, the system transitions to Mode 2 for finer grained localization using spectrum level data. 

\noindent
\textbf{$\blacksquare$ Mode 2: Incumbent Radar Localization Using Spectrograms:}
Upon detection of radar presence in Mode~1, the xApp transitions to Mode~2 by issuing a command to the RAN to begin forwarding I/Q samples in addition to the ongoing KPM collection. The observation window for I/Q data is also set to $t_s$, during which $N$ I/Q samples are gathered and processed.

These raw samples are transformed into two-dimensional spectrograms via the Short-Time Fourier Transform (STFT), providing a joint time–frequency representation of the received signals. The spectrograms offer a detailed view of spectral activity, capturing both cellular and radar transmissions, even under noisy or interference-prone conditions.

For radar signal localization, we adopt the medium variant of the YOLOv11 model (YOLOv11m), which offers a strong balance between localization accuracy and real-time inference capability. YOLO-based architectures have been widely used for spectrogram-based signal detection, including in works such as~\cite{reus2023senseoran, soltani2022finding, ghosh2024sparc}, which demonstrate their suitability for radar pulse detection in the CBRS band.

In our implementation, YOLOv11m processes the spectrograms to identify the precise time–frequency regions occupied by incumbent radar pulses. The model output provides bounding boxes around active radar components, enabling the system to estimate the PRBs affected by radar interference when passed to the  \textit{Dynamic spectrum control module}. It should be noted that once the radar becomes absent, with the aid of the constant KPMs collection, the system automatically switches back to mode 1.

The details on the dataset and evaluations for both modes are provided in Section \ref{datagen} and Section \ref{eval} respectively

\textit{\textbf{Key Observation.}} Note that, unlike previous work in literature ~\cite{reus2023senseoran}, which relies solely on spectrograms for radar detection (and does not implement radar localization or spectrum control mechanism), O-DSS leverages a hybrid mechanism: KPMs for radar detection and spectrograms for radar localization. This design choice is motivated by the minimal data collection overhead of KPMs, which is only kilobytes (KBs) of data compared to the megabytes (MBs) or even gigabytes (GBs) in case of spectrograms. As a result, O-DSS enables significantly faster radar detection; achieving radar detection latency of $\sim$60 ms as seen in our evaluation in Section \ref{e2etiming}, in contrast to nearly 900 ms reported in \cite{reus2023senseoran}. 
Additionally, this strategy reduces capacity overhead on the E2 interface, facilitating better coexistence with other xApps responsible for diverse RAN functionalities.

\vspace{-0.05in}
\subsection{Dynamic Spectrum Control Module}
\vspace{-0.05in}
Based on the inference results from the radar detection and localization module, the \textit{Dynamic spectrum control module} dynamically executes appropriate RAN reconfiguration strategies. These spectrum control policies are categorized into two classes: \textit{coexisting} and \textit{non-coexisting}. This paper primarily focuses on \textit{coexisting policies}, which enables harmonious spectrum sharing while minimizing disruptions to both incumbent radar systems and cellular network performance. Below, we describe two key coexistence strategies implemented in our framework.

\noindent
$\blacksquare$ \textbf{PRB Blanking:}
PRB blanking is employed to mitigate interference from cellular transmissions to radar systems. In this strategy, PRBs overlapping with the radar’s active spectral regions are selectively deactivated at the base station. Prior studies~\cite{lacava2025dapps, ghosh2024sparc} have shown the effectiveness of this approach in protecting incumbent systems while preserving cellular performance over unblanked PRBs. Radar signals, particularly those with pulsed characteristics governed by the Pulse Repetition Rate (PRR), tend to occupy time–frequency subspaces intermittently. Using bounding box outputs from the YOLO-based radar localization module (Mode~2), the original spectrogram dimensions, channel bandwidth and number of PRBs, we are able to estimate the PRBs utilized for radar transmission. These PRBs are then blanked, while unaffected PRBs remain active. When no radar is detected, all PRBs are re-enabled, optimizing spectrum utilization.

\noindent
$\blacksquare$ \textbf{Radar-Aware MCS Optimization:}
While PRB blanking mitigates main-lobe interference, it cannot eliminate residual interference caused by radar side lobes. For instance, radar types such as P0N \#1 and P0N \#2 (considered in this paper) emit significant energy into adjacent frequencies and directions, impacting PRBs outside the blanked region. This persistent impairment justifies claims in~\cite{sanders2013radar} that P0N \#2 causes the greatest degradation in cellular performance. To address this, we implement a \textit{radar-aware MCS optimization algorithm}, shown in Algorithm~\ref{mcs_Adapt}. This algorithm adaptively tunes the MCS based on observed trends in the BLER, complementing Channel Quality Indicator (CQI)-based adaptation with direct feedback on link-layer performance. Conventional RAN MCS selection primarily relies on CQI reports, which estimate the SINR conditions observed at the UE. However, CQI reporting may not adequately capture the effects of transient or non-Gaussian interference such as radar main or side lobes especially when such interference is bursty or sporadic. Unlike CQI, BLER provides a post-transmission performance metric, directly reflecting the success or failure of packet transmissions. By incorporating BLER into the MCS decision logic, we introduce a feedback-driven mechanism that allows the base station to react to actual link-layer impairments rather than predicted channel quality alone.

The MCS optimization logic adopts a modified Additive Increase and Multiplicative Decrease (AIMD) strategy similar to those used in popular congestion control algorithms like TCP Reno with provable stability guarantees~\cite{jacobson1988congestion, grieco2004performance}. Unlike these traditional algorithms, our algorithm does not employ a slow-start phase and instead initializes with the MCS index received from the RAN. The system then adaptively tunes the MCS index based on BLER feedback: decreasing it multiplicatively when BLER exceeds $\text{BLER}_{\text{thresh}}$, increasing it additively when BLER is low, and holding it constant when BLER remains within an acceptable range ($\gamma$). This approach balances robustness and spectral efficiency under rapidly changing interference and link quality conditions.

Alternative coexistence strategies such as transmit/receive power control and Inter-cell Interference Coordination (ICIC) are left for future extensions of the framework.

\begin{algorithm}[!ht]
\small
\caption{Radar-Aware MCS Optimization Algorithm}
\begin{algorithmic}[1]
\Require $MCS_{BS}$ (current MCS index at BS), $BLER$ (current BLER observed at BS), $BLER_{prev}$ (previous BLER), $A$ (action), $\gamma$ (BLER variation threshold), $BLER_{thresh}$ (maximum acceptable BLER), $MCS_{max}$, $MCS_{min}$
\Ensure Updated value of $MCS$, the MCS index at BS
\If{$|BLER - BLER_{prev}| < \gamma$}
    \State \Return $MCS_{BS}$ \Comment{No significant change in BLER}
\EndIf
\If{$BLER > BLER_{thresh}$}
    \State $A \gets DECR$
    \State $MCS \gets \max(MCS_{BS} // \beta, MCS_{min})$
\Else
    \State $A \gets INCR$
    \State $MCS \gets \min(MCS_{BS} + \beta, MCS_{max})$
\EndIf
\State \Return $MCS_{BS}$
\end{algorithmic}
\label{mcs_Adapt}
\end{algorithm}
\vspace{-0.2in}
\section{O-DSS System Implementation}
\label{systemimplementation}
\vspace{-0.02in}
\subsection{System Overview}
\vspace{-0.02in}
To prototype O-DSS system, we begin with a scenario involving shipborne radar systems, and later extend our O-RAN system to support airborne radar systems as discussed in Section~\ref{aerial_radar}. We employ a hybrid setup that combines Commercial-Off-The-Shelf (COTS) hardware and Software-Defined Radio (SDR) components for Over-The-Air (OTA) evaluations, along with a virtual RF simulation environment for extensive, scalable experimentation.

\begin{figure}[!t]
    \centering
    \includegraphics[width=0.5\textwidth]{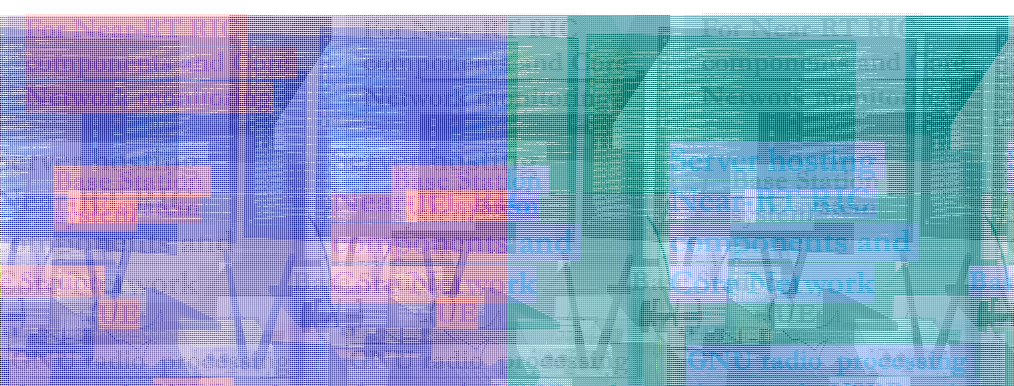}
        \caption{OTA testbed implementation of the O-DSS framework. The setup includes: (i) GNU Radio host processing stored radar waveforms for transmission; (ii) USRP transmitting radar signals in-band; (iii) BS and UE running on \texttt{srsRAN} based OAIC stack; (iv) monitoring stations for BS, UE, Near-RT RIC and core network components; and (v) a server rack hosting the containerized Near-RT RIC components/modules.}

    \label{fig:testbed}
    \vspace{-0.2in}
\end{figure}

Figure~\ref{fig:testbed} illustrates our OTA testbed. The cellular base station and the User Equipment (UE) are implemented using the open-source \texttt{srsRAN} 4G/5G stack. Specifically, we adopt the Open AI Cellular (OAIC) platform~\cite{oaic}, which builds on \texttt{srsRAN} v21.10~\cite{srs} to enable AI-native experimentation.

The base station and UE run on Ubuntu 22.04 desktops powered by Intel Core i9-14900K processors (32 logical cores, 64GB RAM, up to 6GHz). Experiments are conducted in Frequency Division Duplex (FDD) mode over a 10MHz system bandwidth (50 PRBs) using 3GPP Band 7: 2.68GHz downlink and 2.56GHz uplink. Uplink traffic, generated using \texttt{iperf3}, is our primary focus for performance analysis. Both KPMs and I/Q samples are collected at the base station. While the OTA experiments involve a single UE, the results generalize to RAN-wide behavior as our interest lies in aggregate network performance. In the TDD case, the sensing logic would adapt to monitor specific uplink slots or symbols, similar to~\cite{lacava2025dapps}.

Radar signals are generated offline and transmitted using a third USRP device running a GNU Radio pipeline. The radar transmitter emits signals in the same uplink band as cellular traffic, emulating coexistence. Detailed waveform generation parameters are described in Section~\ref{datagen}.

To evaluate robustness beyond physical hardware, we also leverage the ZMQ-based virtual RF interface in \texttt{srsRAN}~\cite{srsran_zeromq}. This RF simulation backend supports multi-UE prototyping, traffic control, and SINR manipulation via GNU Radio, enabling controlled spectrum sharing studies at scale.

The Near-RT RIC is hosted on a server with Intel Xeon Gold 6526Y CPUs and 64GB RAM. Its components including the policy controller, E2 interface, and the O-DSS xApp are deployed as modular Docker containers for scalability. Communication between the gNB and RIC occurs over the E2 interface using SCTP, following the high-level architecture shown in Figure~\ref{fig:odss}.

\begin{figure}[!b]
\vspace{-0.1in}
    \centering
    \begin{subfigure}[b]{0.15\textwidth}
        \centering
        \includegraphics[width=\linewidth]{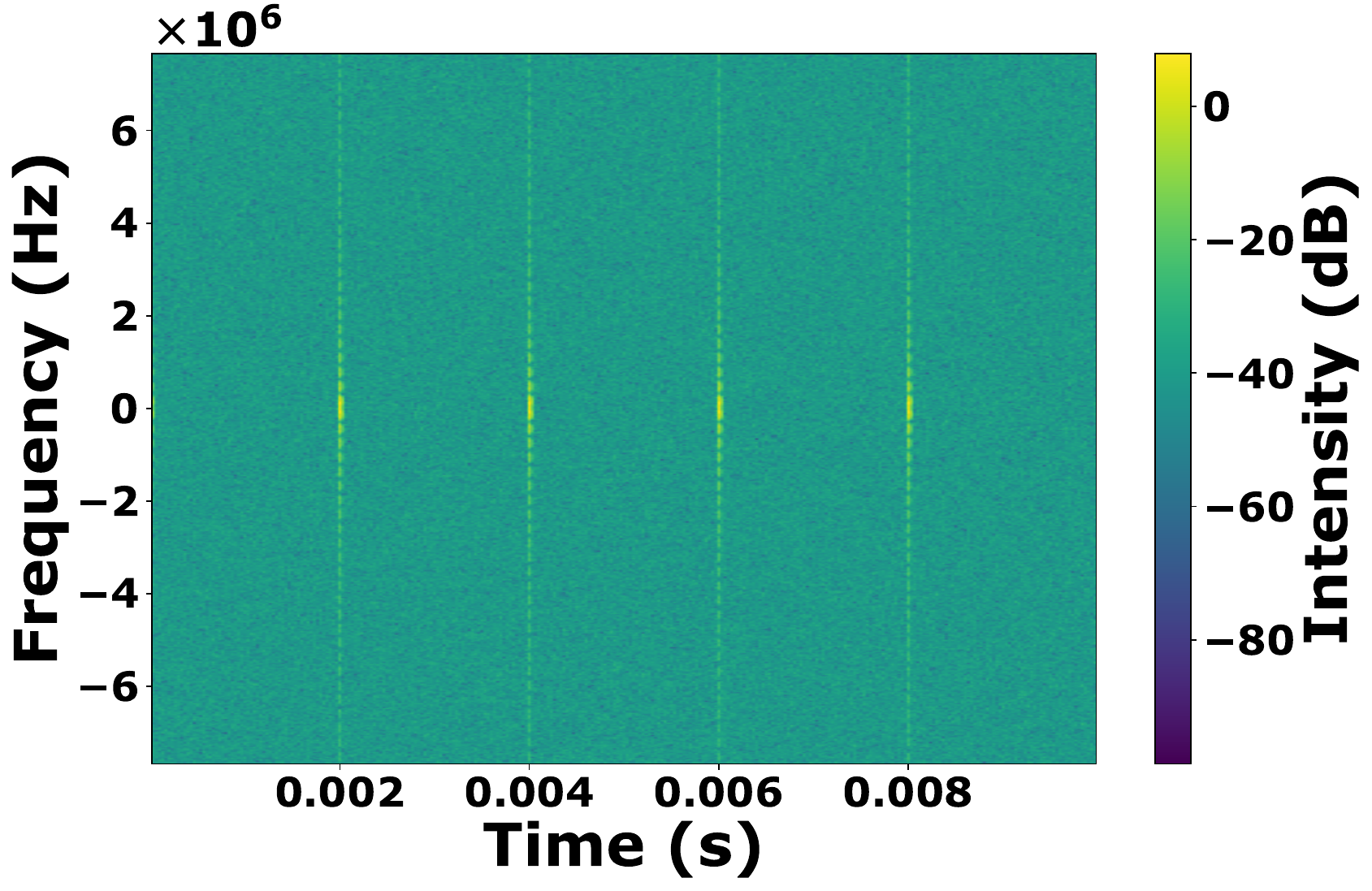}
        
        \caption{Radar signal at 10 dB SNR}
        
        \label{fig:radar10}
    \end{subfigure}
    \hfill
    \begin{subfigure}[b]{0.15\textwidth}
        \centering
        \includegraphics[width=\linewidth]{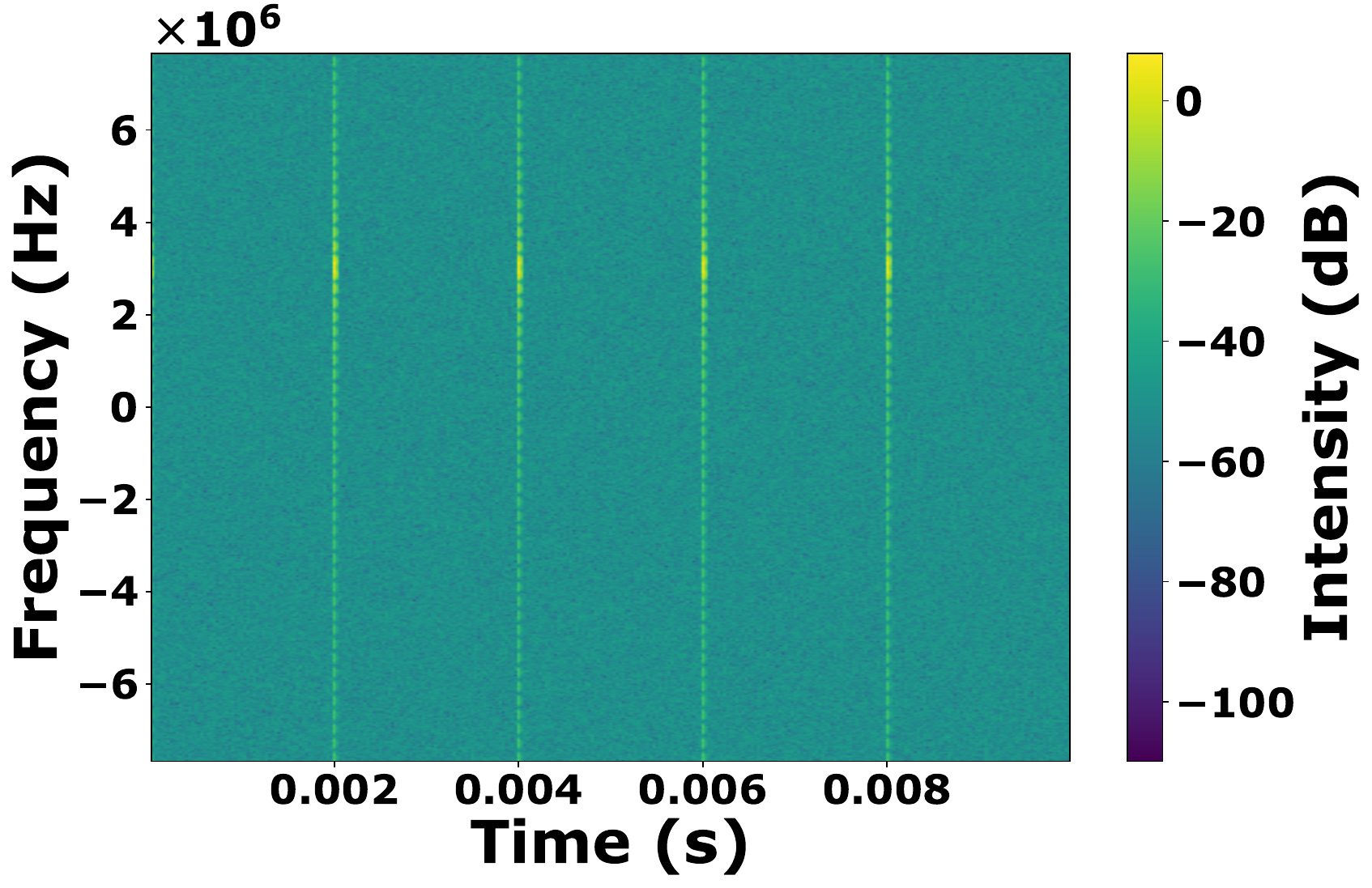}
        \caption{Radar signal at 20 dB SNR}
        \label{fig:radar20}
    \end{subfigure}
    \hfill
    \begin{subfigure}[b]{0.15\textwidth}
        \centering
        \includegraphics[width=\linewidth]{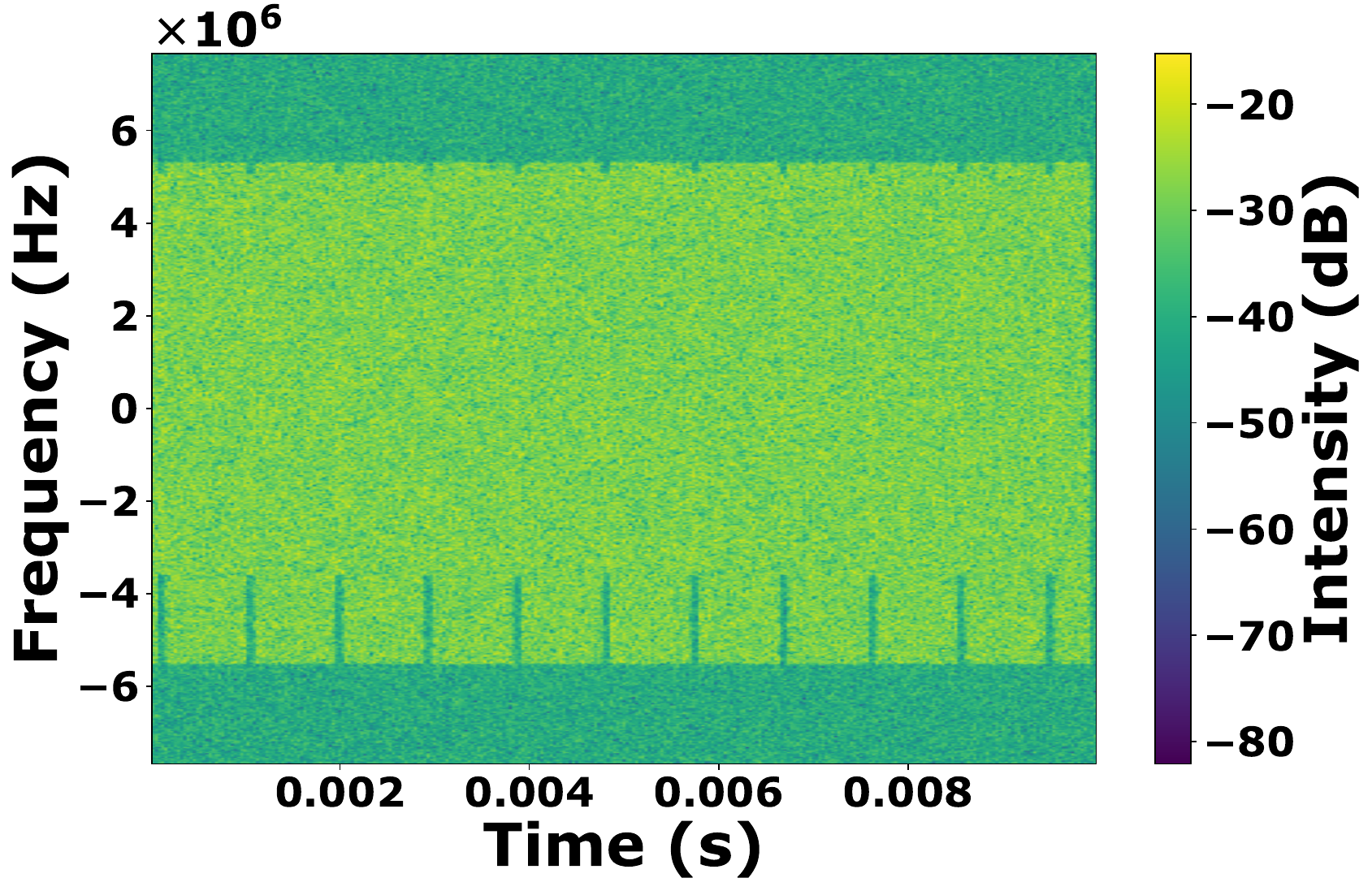}
        \caption{Cellular signal at 10 dB INR}
        \label{fig:lte10}
    \end{subfigure}
    \caption{Spectrograms of radar and cellular signals under varying Signal-to-Noise (SNR) and Interference-to-Noise (INR) conditions. Radar signal having pulse width of 13 $\mu$s, PRR of 500, pulses per burst is 5 and burst length is 10 ms.}
    \label{fig:combined-spectrograms}
    \vspace{-0.2in}
\end{figure}
\vspace{-0.04in}
\subsection{Dataset Generation}
\label{datagen}

\begin{figure*}[!t]
    \centering
    \begin{subfigure}[b]{0.45\textwidth}
        \centering
        \includegraphics[width=\linewidth]{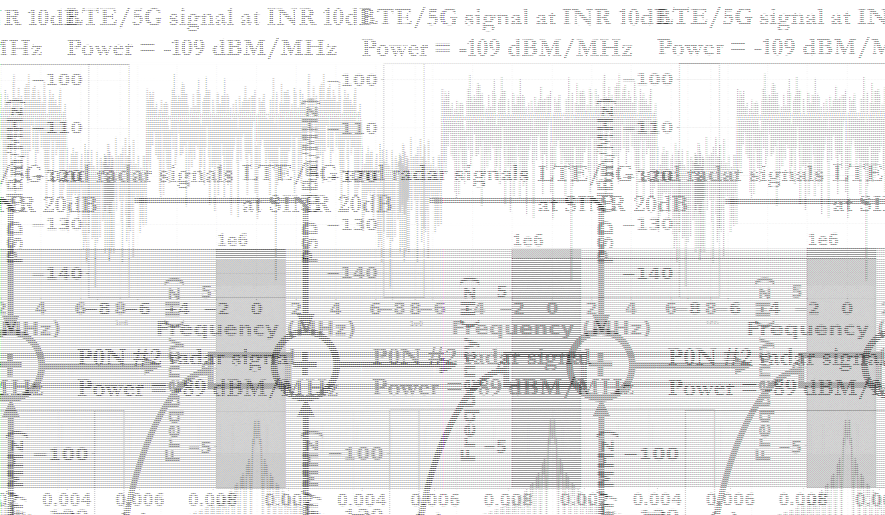}
        \caption{Radar + cellular signal PSD.}
        \label{fig:radar-lte}
    \end{subfigure}
    \hfill
    \begin{subfigure}[b]{0.45\textwidth}
        \centering
        \includegraphics[width=\linewidth]{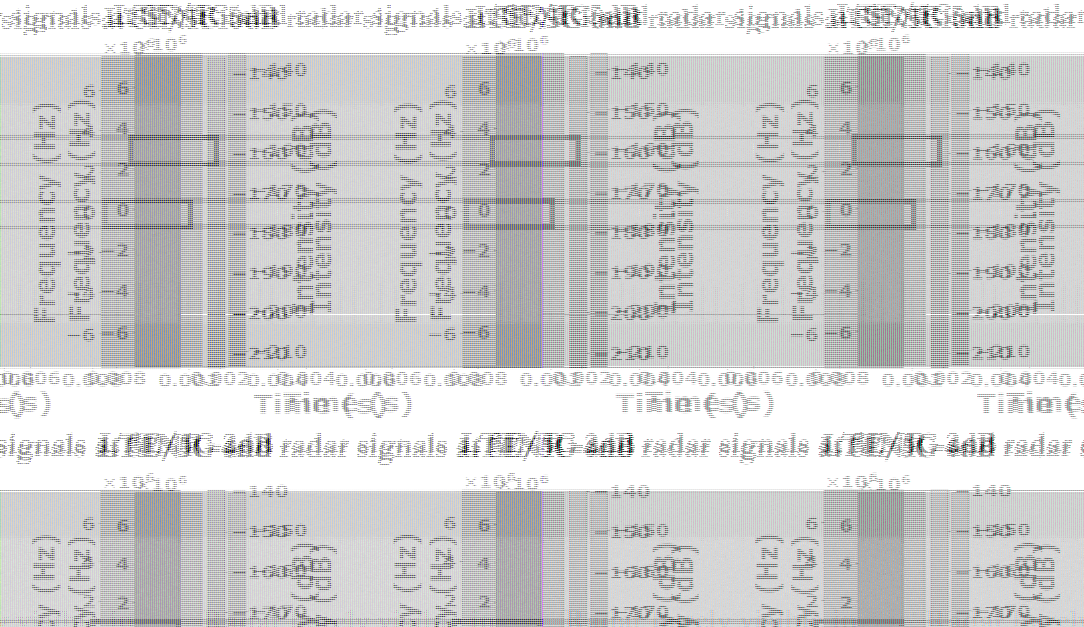}
        \caption{Radar + cellular signal PSD under varying SINR.}
        \label{fig:radar-lte-varying}
    \end{subfigure}
    \vspace{-0.1in}
    \caption{Power spectral densities of radar and cellular signals under different SINR conditions.}
    \label{fig:combined-radar-lte}
    \vspace{-0.2in}
\end{figure*}
\vspace{-0.04in}
Both KPMs and spectrogram datasets used for training the models in the radar detection and localization module are generated through the fusion of cellular and radar waveforms. Radar signal characteristics follow standard compliant parameters outlined by NIST \cite{cbrs_dataset}. We focus on Type 2 radar signals, characterized by fixed-frequency carrier wave pulse modulation (P0N \#2). This radar type exhibits one of the narrowest spectral footprints among those evaluated, making it ideal for analyzing radar–cellular coexistence. Other radar types, such as Q3N \#1–\#3, which employ swept-frequency pulse modulation and occupy significantly more bandwidth, are better suited for non-coexistence scenarios and are reserved for future study. Prior analysis in \cite{sanders2013radar} shows that P0N \#2 signals introduce considerable degradation in cellular throughput, further motivating their selection.

Radar waveforms are generated using MATLAB and the NIST tool \cite{caromi2020simulated}, allowing control over parameters such as pulse width (13–52 $\mu$s), PRR (500–1100 Hz), and pulses per burst (15–40). Generated I/Q samples are sampled at $f_{\mathrm{s}} = 15.36$ Msps to match the sampling rate of the cellular system. Unlike ESC sensors that monitor the full 100 MHz CBRS band, our experiments focus on a 10 MHz slice of spectrum corresponding to the operating bandwidth of the base station. We set  10 ms for the observation window (see  Figure \ref{fig:combined-spectrograms}).

Initial spectrograms are created via offline DSP to ensure precise SINR control in accordance with FCC regulations. Radar signal power is treated as the desired signal, while cellular signal power and AWGN serve as the interference and noise components. SINR is computed as

\begin{equation}
\text{SINR} = 10 \times \log_{10} \left( \frac{P_{\text{radar}}}{P_{\text{cellular}} + P_{\text{noise}}} \right),
\label{eq:sinr_radar}
\end{equation}
where $P_{\text{radar}}$, $P_{\text{cellular}}$, and $P_{\text{noise}}$ represent the radar, interference, and noise power levels, respectively, measured in dBm/MHz. Per FCC requirements, cumulative interference and noise must not exceed -109 dBm/MHz, and radar pulses with at least -89 dBm/MHz (equivalent to 20 dB SINR) must be detected with 99\% accuracy \cite{sanders2017procedures}. To go beyond current detection limits, we also generate data at SINR levels between -4 and 12 dB. This is achieved by fixing interference and noise at -109 dBm/MHz and varying radar power from -105 to -97 dBm/MHz. To enhance diversity, radar pulses are inserted at three distinct center frequencies overlapping with the cellular band. Figure \ref{fig:combined-radar-lte} illustrates this setup.

Additional spectrograms are generated by playing back the radar I/Q samples through GNU Radio and transmitting them via ZMQ RF simulator or a USRP (OTA) using the setup shown in Figure \ref{fig:testbed}. KPMs are collected from these environments at varying SINR levels by adjusting radar and cellular node placements or transmit power. Uplink traffic, ranging between 1–5 Mbps, is generated using \texttt{iperf3}.

We generated spectrograms for three scenarios: (i) cellular + noise only, (ii) radar + cellular + noise at different SINRs, and (iii) radar centered at various frequencies. We collected KPMs under analogous conditions and included throughput, BLER, MCS, SINR, and BSR. We generated over $14,000+$ spectrogram samples and more than $10,000+$ KPM observations through a combination of offline signal processing, emulation, and OTA experiments.

\vspace{-0.05in}
\section{Experimental Evaluation}
\label{eval}
\vspace{-0.04in}
This section presents a comprehensive evaluation of the prototyped O-DSS framework (See Section IV) that enables coexistence of cellular networks in CBRS band with shipborne radar systems as primary incumbents. We begin by analyzing the performance of the radar detection and localization modules of the O-DSS framework. Next, we demonstrate the framework using both the RF simulator and OTA testbed to assess its impact on cellular performance. Finally, we evaluate the end-to-end system latency to verify compliance with O-RAN’s Near-RT closed-loop control constraints.

\vspace{-0.02in}
\subsection{Radar Detection Evaluation}
\label{detect_eval}
\begin{figure}[!t]
    \centering
    \includegraphics[width=0.5\textwidth]{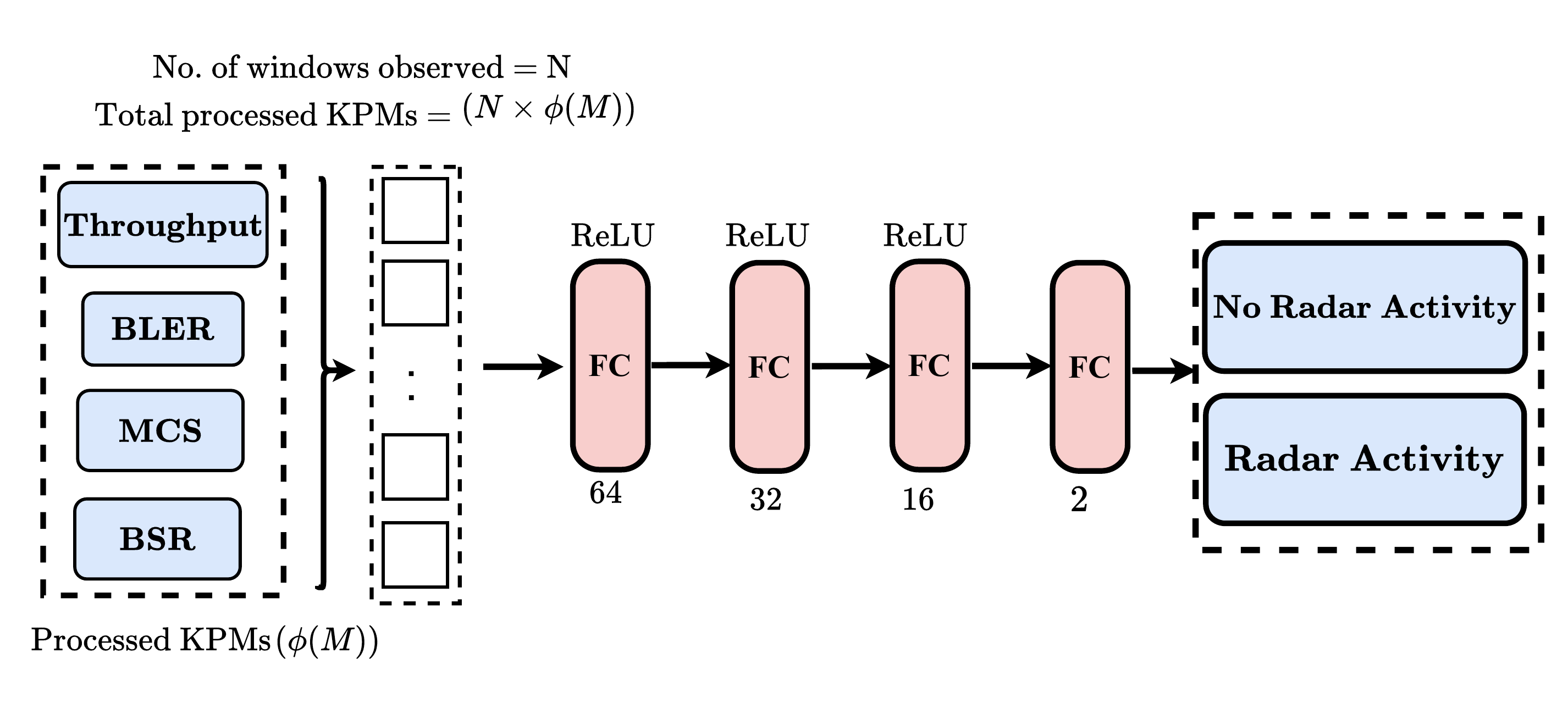}
    \vspace{-0.2in}
        \caption{DNN model architecture for incumbent radar activity detection using processed RAN KPMs.}
    \label{fig:kpm_model}
    \vspace{-0.25in}
\end{figure}

\vspace{-0.02in}
To detect incumbent radar activity, we design a lightweight DNN that ingests RAN KPMs such as throughput, BLER, MCS, and BSR collected over a sliding time window. As illustrated in Figure ~\ref{fig:kpm_model}, the DNN comprises a stack of ReLU-activated fully connected layers, enabling it to learn nonlinear temporal dependencies in the input sequence.

The model contains only $3{,}474$ trainable parameters, making it suitable for near-real-time inference on resource-constrained RAN nodes. Training is performed on 75\% of the $10{,}000+$ KPM dataset (see Section~\ref{datagen}), with normalization and shuffling to ensure generalization. The input tensor is shaped as $(N \times M, 1)$, where $M = 4$ is the number of features and $N$ the sliding window size. We use the RMSprop optimizer (learning rate = 0.001) and sparse categorical cross-entropy loss, training over 50 epochs with batch size 128.

To benchmark performance, we compare this DNN to an eXtreme Gradient Boosting (XGBoost) model, a high-performing ensemble method tailored to structured data. The \texttt{XGBClassifier} is configured for multiclass classification (\texttt{multi:softmax}) with hyperparameters: maximum depth = 4, 100 boosting rounds, and learning rate = 0.1.

We evaluate both models across four sliding window lengths ($N = \{1, 2, 3, 4\}$) to quantify the impact of temporal aggregation. As shown in Figure ~\ref{fig:dnn_xgboost}, increasing $N$ consistently improves classification accuracy, validating the benefit of temporal context. Both models deliver comparable performance; notably, the DNN achieves accuracy on par with that of XGBoost across all window sizes $\geq 2$.

Finally, on testing, the trained KPM-based DNN achieves 100\% radar detection accuracy for SINR $\geq -4\,$dB. Below this threshold, radar interference no longer degrades cellular performance, rendering detection impossible using KPMs.

\vspace{-0.05in}
\subsection{Radar Localization Evaluation}
\label{radar_localization}
\vspace{-0.05in}
We now evaluate the radar localization component of the O-DSS's incumbent detection and localization module. Complementing the KPM-based detection model, this component employs a custom YOLOv11m model to identify radar emissions in the frequency domain using spectrogram inputs. 

The model is trained on a custom dataset of over $14{,}000+$ radar-overlaid spectrograms (see Section~\ref{datagen}), preprocessed to fit the model’s input dimensions. Training is performed over 100 epochs with a batch size of 64 using Stochastic Gradient Descent (SGD). A grid search across learning rates \{0.1, 0.01, 0.001\} determined 0.01 to yield the best convergence-stability tradeoff. The loss function includes classification loss (signal type), bounding box regression loss, and Distribution Focal Loss for improved boundary precision.

\begin{figure}[!t]
    \centering
    \begin{subfigure}[b]{0.49\linewidth}
        \centering
        \includegraphics[width=\linewidth]{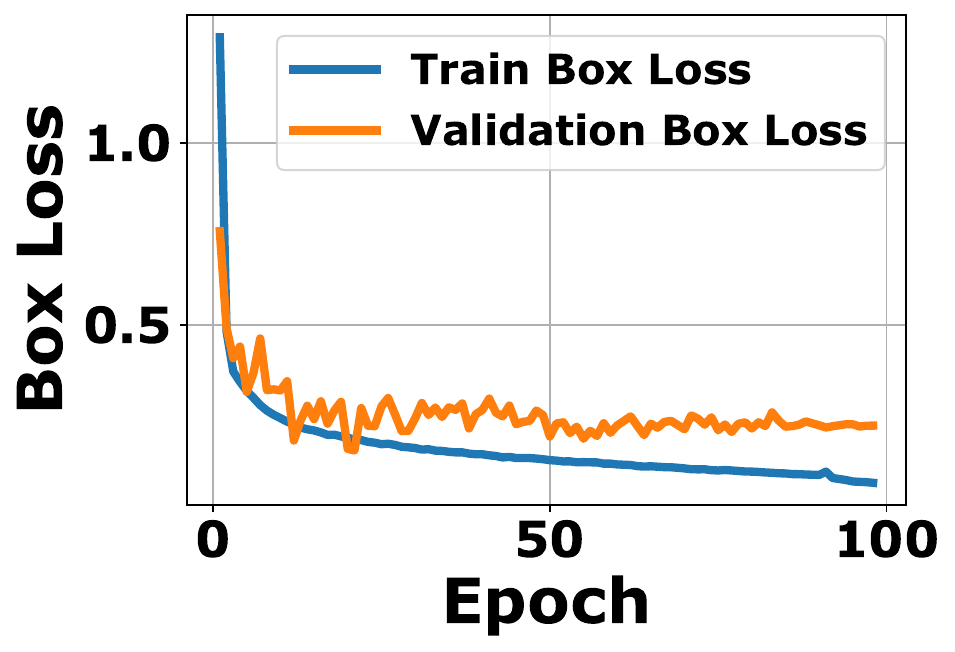}
        \caption{Training and validation box loss across epochs.}
        \label{fig:boxloss}
    \end{subfigure}
    \hfill
    \begin{subfigure}[b]{0.49\linewidth}
        \centering
        \includegraphics[width=\linewidth]{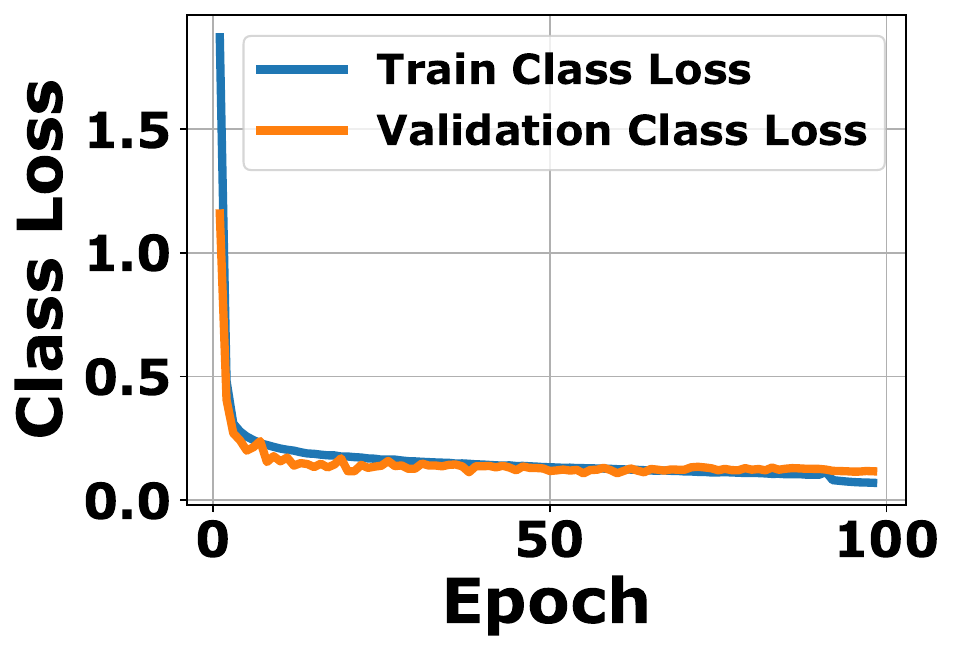}
        \caption{Training and validation classification loss across epochs.}
        \label{fig:clsloss}
    \end{subfigure}

    \vspace{-0.07in}
    \caption{Custom YOLOv11 model training performance, showing loss convergence for bounding box regression and classification.}
    \label{fig:radarlocalization}
    \vspace{-0.25in}
\end{figure}

\begin{figure}[!b]
\vspace{-0.2in}
    \centering
    \begin{subfigure}[b]{0.49\linewidth}
        \centering
        \includegraphics[width=\linewidth]{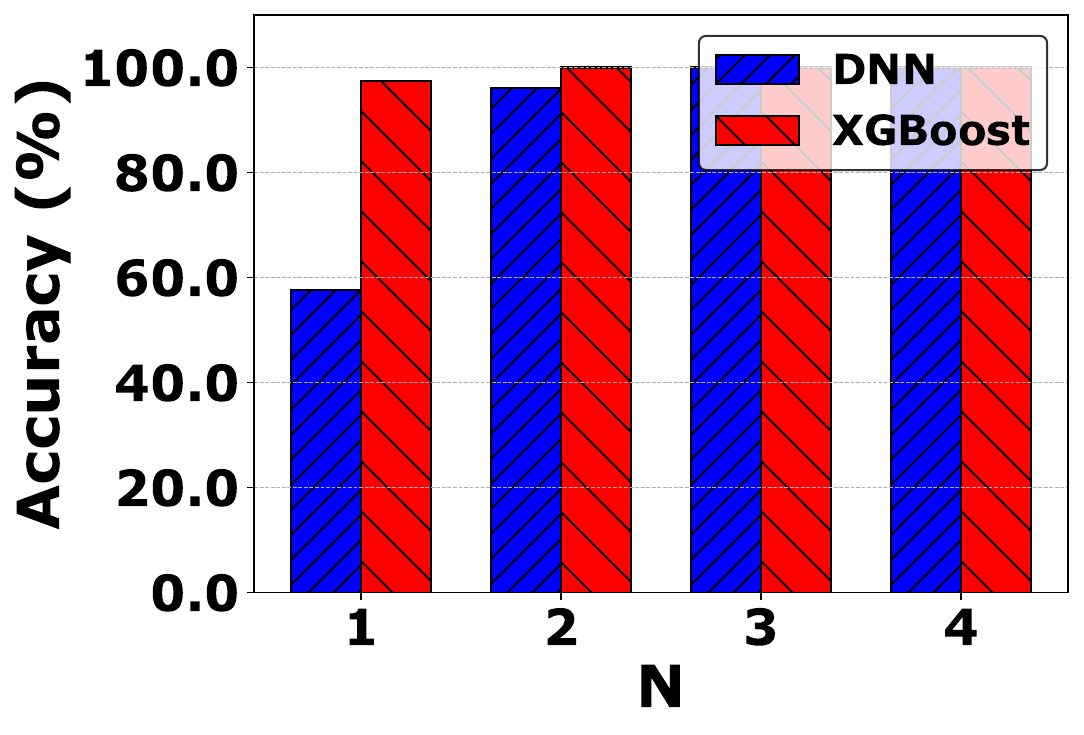}
        \caption{DNN vs XGBoost accuracy for different window sizes $N$.}
        \label{fig:dnn_xgboost}
    \end{subfigure}
    \hfill
    \begin{subfigure}[b]{0.49\linewidth}
        \centering
        \includegraphics[width=\linewidth]{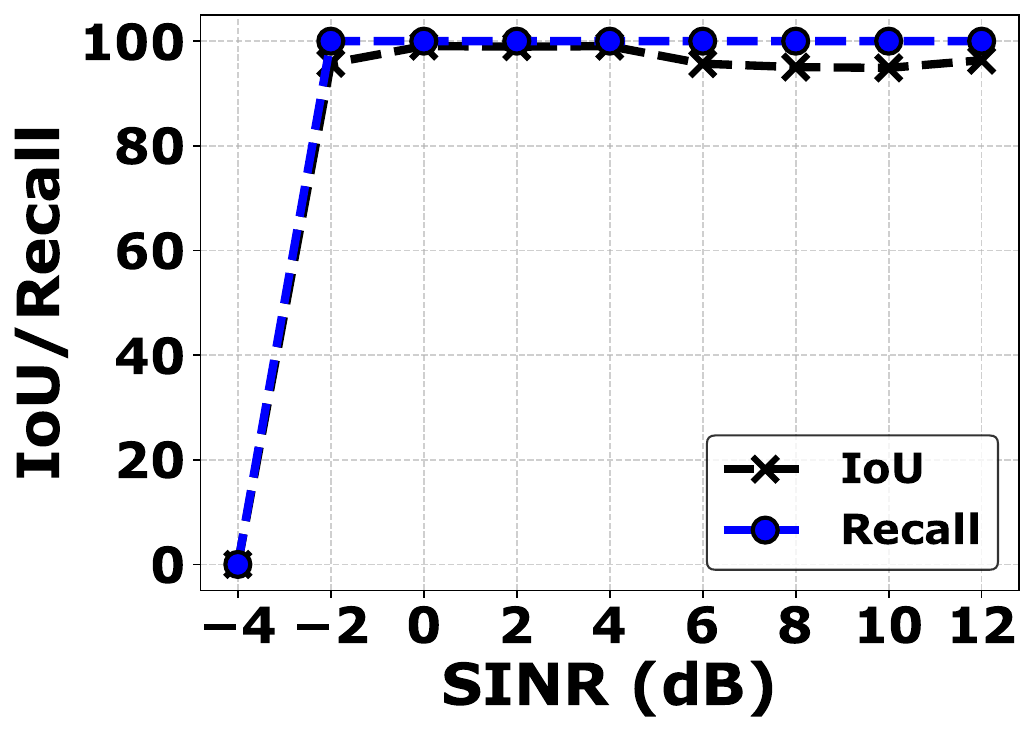}
        \caption{IoU and Recall under SINR conditions.}
        \label{fig:iourecall}
    \end{subfigure}
    \vspace{-0.2in}
    \caption{(a) DNN-based radar detection: Increasing temporal observation window improves accuracy for both models. (b) Custom YOLO model based radar localization performance on spectrograms under varying SINR.}
    \label{fig:model_metrics}
\end{figure}

Figures~\ref{fig:boxloss} and~\ref{fig:clsloss} show steady convergence of training and validation losses, indicating robust generalization. Performance across varying SINR levels is shown in Figure~\ref{fig:iourecall}, where Intersection over Union (IoU) exceeds $95\%$ and Recall surpasses $99\%$ beyond $0$~dB, confirming high reliability in moderate to high SINR regimes. Degradation at negative SINRs ($\leq-4$) highlights the need for detection strategies under lower-SINR conditions.

Given a spectrogram input of shape \((N \times M)\), where \(N\) and \(M\) correspond to frequency and time axes respectively, the YOLOv11m model outputs bounding boxes for both radar and cellular signal regions. Specifically, the model maps the input \(\mathcal{I}^{N \times M} \rightarrow \mathcal{I}[f_L, f_H, F_L, F_H]\), where \([f_L, f_H]\) and \([F_L, F_H]\) denote the frequency bounds of the radar and cellular signals, respectively. The radar bounding box \([f_L, f_H]\) is used by the PRB blanking module to determine which PRBs are impacted by incumbent radar activity, thereby ensuring regulatory compliance and minimizing degradation in cellular performance.

\begin{figure}[!t]
    \centering
    \includegraphics[width=0.5\textwidth]{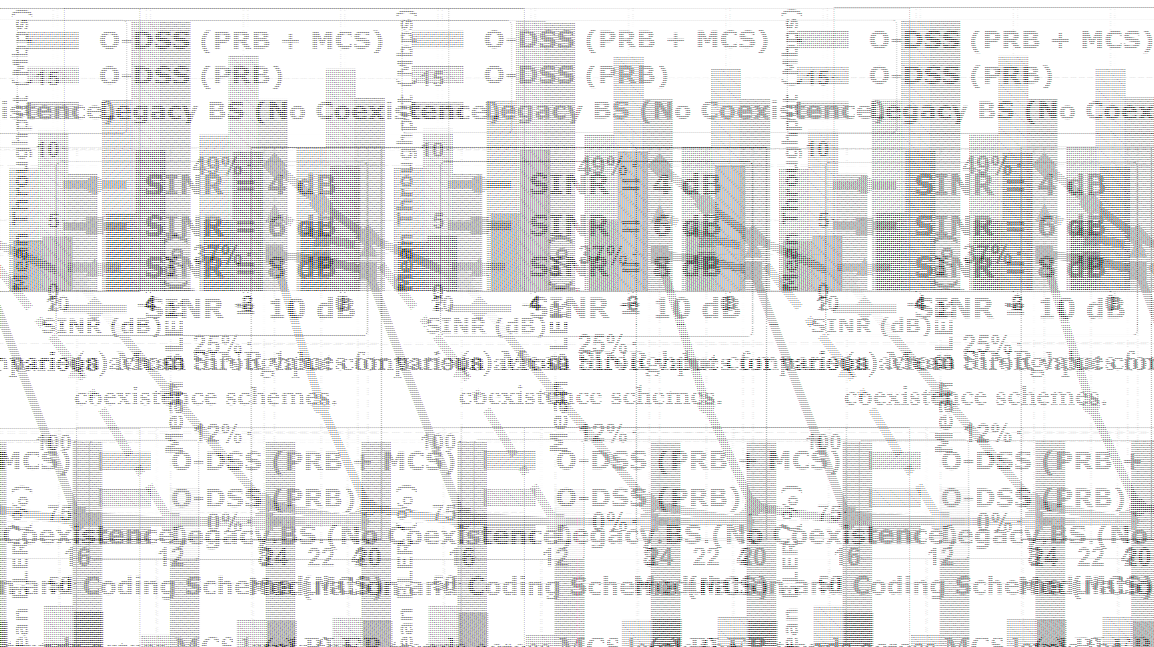}
    \caption{Performance comparison of baseline and O-DSS spectrum coexistence strategies under radar transmission: (a) average uplink throughput vs SINR, (b) average BLER vs SINR, and (c) mean BLER across MCS levels at selected SINR values.}
    \label{fig:zmq_performance}
    \vspace{-0.2in}
\end{figure}

\begin{figure}[!b]
\vspace{-0.2in}
    \centering
    \includegraphics[width=0.5\textwidth]{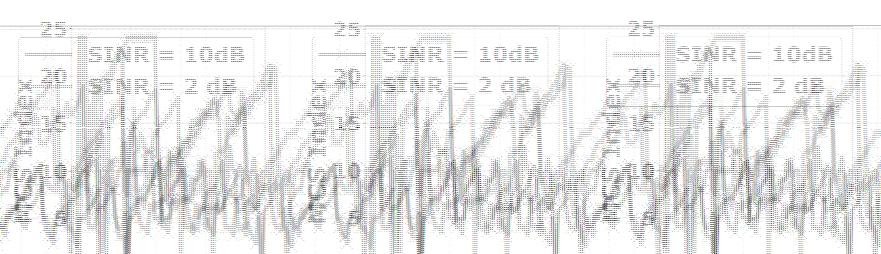}
           \caption{MCS index adaptation using Algorithm \ref{mcs_Adapt} with $\gamma = 1$ and $\beta = 2$, starting from the maximum MCS (received from RAN). The additive increase and multiplicative decrease balance throughput and BLER under observed SINR values.}
    \label{fig:mcs_time}
    \vspace{-0.1in}
\end{figure}

\vspace{-0.07in}
\subsection{Cellular Network Performance Evaluation}
\vspace{-0.07in}
We evaluate the proposed O-DSS framework under radar transmission using both the ZMQ-based RF simulator and the OTA USRP testbed (Section~\ref{systemimplementation}). For each configuration, we measure uplink throughput and BLER at the base station across SINR values from \(-4\) to \(10\) dB, using full-bandwidth (10 MHz) uplink traffic from all UEs.

Figures~\ref{fig:zmq_performance}(a) and (b) compare three schemes: (i) baseline (no coexistence mechanism), (ii) O-DSS (PRB blanking only), and (iii) Full O-DSS (PRB blanking + radar-aware MCS). For the full O-DSS system, we apply Algorithm \ref{mcs_Adapt}, fixing $BLER_{thresh} = 5\%$, $\gamma = 1$ and tuning $\beta$ through extensive experiments. We then select $\beta = 2$ as it minimizes BLER while maintaining acceptable throughput.

At \(-4\) dB SINR, PRB blanking alone offers limited gains due to poor spectrogram-based radar localization, but KPM-based detection enables robust MCS selection. This significantly reduces BLER, intentionally trading throughput for reliability. As SINR increases, the baseline system rapidly degrades, while O-DSS variants sustain robust performance.

Figure~\ref{fig:zmq_performance}(c) shows BLER across MCS levels for different SINRs. Persistently high BLER at higher MCS even with PRB blanking clearly motivates the need for radar-aware MCS tuning. The full O-DSS stack effectively adapts to radar conditions, enabling reliable and efficient coexistence.

Figure~\ref{fig:mcs_time} demonstrates how Algorithm \ref{mcs_Adapt} responds to different SINR conditions. At lower SINR of 2 dB, the algorithm maintains a higher average MCS compared to the 10 dB scenario. This is attributed to lower and more stable BLER trends at low SINR, enabling steady additive increases, whereas higher SINR suffers abrupt link fluctuations triggering frequent multiplicative reductions and lower MCS indexes.

\begin{figure}[!t]
    \centering
    \includegraphics[width=0.5\textwidth]{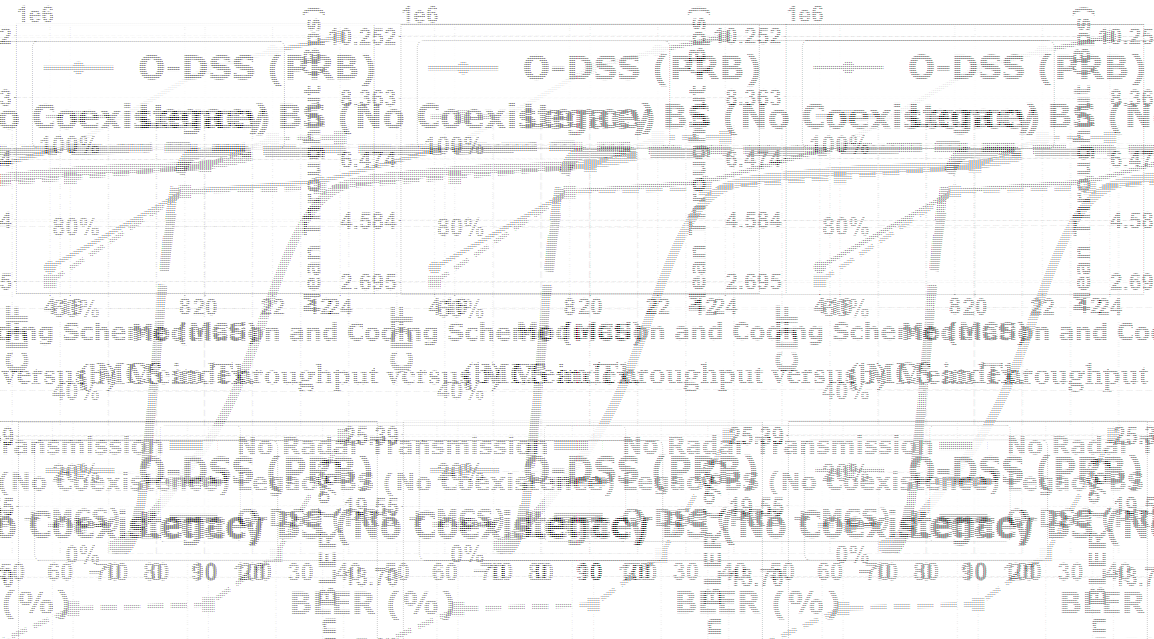}
    \caption{Performance evaluation of O-DSS on O-RAN testbed showing (a) cumulative BLER distribution, (b) UL throughput, and (c) UL BLER across MCS values.}

    \label{fig:ota_performance}
    \vspace{-0.2in}
\end{figure}

Finally, we assess the OTA performance of the O-DSS framework using the USRP testbed. Figure~\ref{fig:ota_performance}(a) shows that the full O-DSS stack combining PRB blanking with radar-aware MCS adaptation closely tracks the radar-free baseline, significantly outperforming the legacy system. Figures~\ref{fig:ota_performance}(b) and (c) further underscore the critical role of radar-aware MCS adaptation: while PRB blanking alone enables moderate gains, its integration with dynamic MCS control ensures robust operation. This validates that joint spectrum and link-level control is essential for resilient cellular–radar coexistence.

\vspace{-0.1in}
\subsection{O-DSS Framework Timing Evaluation}
\label{e2etiming}
\vspace{-0.07in}
We conduct a comprehensive latency breakdown in Table~\ref{tab:odss_timing} of the O-DSS system framework shown in Figure \ref{fig:odss}. We assess its suitability for Near-RT spectrum control under O-RAN constraints. Our evaluation reveals that radar detection via KPMs completes in $\sim$60 ~ms substantially faster than \cite{reus2023senseoran} $\sim$900 ms and well below the FCC's 60 second threshold. Furthermore, end-to-end channel evacuation using spectrogram based radar localization and spectrum control (via PRB blanking) executes in $\sim$700 ms, outperforming legacy standards that allow up to 5 minutes. These results demonstrate that O-DSS not only meets the 1 second control loop requirement of the Near-RT RIC, but also achieves superior responsiveness compared to baseline systems.

\begin{table}[!h]
\centering
\vspace{-0.1in}
\caption{O-DSS Timing Analysis}
\label{tab:odss_timing}
\renewcommand{\arraystretch}{1.2}
\scriptsize
\begin{tabular}{|p{2.5cm}|p{1cm}|p{2.5cm}|p{1cm}|}
\hline
\textbf{Radar Detection} & \textbf{Time} & \textbf{Radar Localization and Dynamic Spectrum Control} & \textbf{Time} \\
\hline
Receive KPMs, store in DB \circled{1} \& \circled{2} & $\sim$11 ms & Receive I/Q, generate spectrogram, store in DB \circled{1} \& \circled{2} & $\sim$450 ms \\
\hline

Process KPMs, Model inference \& policy generation \circled{3} \& \circled{4} (MODE 1)& $\sim$45 ms & Model inference \& policy generation \circled{3} \& \circled{4} (MODE 2) & $\sim$200 ms \\
\hline

Control decision to RAN \circled{5} \& \circled{6} & $\sim$70 $ \mu $s & Control decision to RAN \circled{5} \& \circled{6} & $\sim$70 $ \mu$s \\
\hline
-  & - & Spectrum Control (PRBs Blanked, MCS adaptation) & $\sim$ 12 ms \\
\hline
\textbf{Total Time} & \textbf{$\sim$60 ms} & \textbf{Total Time} & \textbf{$\sim$700 ms} \\
\hline
\end{tabular}
\end{table}


\vspace{-0.1in}
\section{O-DSS for airborne radars (Case Study)}
\label{aerial_radar}
\vspace{-0.05in}
To extend the applicability of the O-DSS framework, we evaluate its performance for airborne radar scenarios in mid-band frequencies using the 3.5 GHz CBRS band as a case study. This marks the first open study to explore cellular–airborne radar coexistence under realistic mid-band channel conditions using COTS hardware and a commercial-grade channel emulator.
\vspace{-0.1in}
\subsection{Experimental Setup and Dataset}
\begin{figure}[!t]
\vspace{-0.15in}
    \centering
    \includegraphics[width=0.49\textwidth]{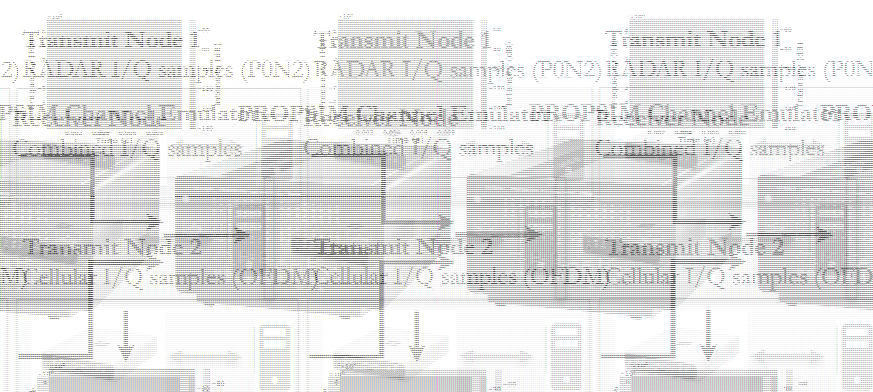}
    \caption{Experimental testbed combining radar and OFDM transmitters connected to a Keysight PROPSIM channel emulator to simulate realistic airborne propagation for mid-band radar–cellular coexistence studies.}
    \label{fig:airborne_testbed}
    \vspace{-0.15in}
\end{figure}

The setup in Figure~\ref{fig:airborne_testbed} combines USRP radios and the Keysight PROPSIM emulator to model high-mobility radar conditions, incorporating altitude, velocity, and multipath effects aligned with 5G/LTE PHY standards. Radar and OFDM signals are transmitted at 15.36 Msps with a 1024 point FFT. Table~\ref{tab:dataset-collection} outlines five scenarios covering Line-of-Sight (LOS), Non-Line-of-Sight (NLOS) paths, varying altitudes, and radar speeds (25–200 m/s), with a fixed OFDM transmitter at 180 m to emulate cellular transmission.

\vspace{-0.05in}
\subsection{Validation of Doppler Effects}
\begin{figure}[!t]
    \centering
    \includegraphics[width=0.49\textwidth]{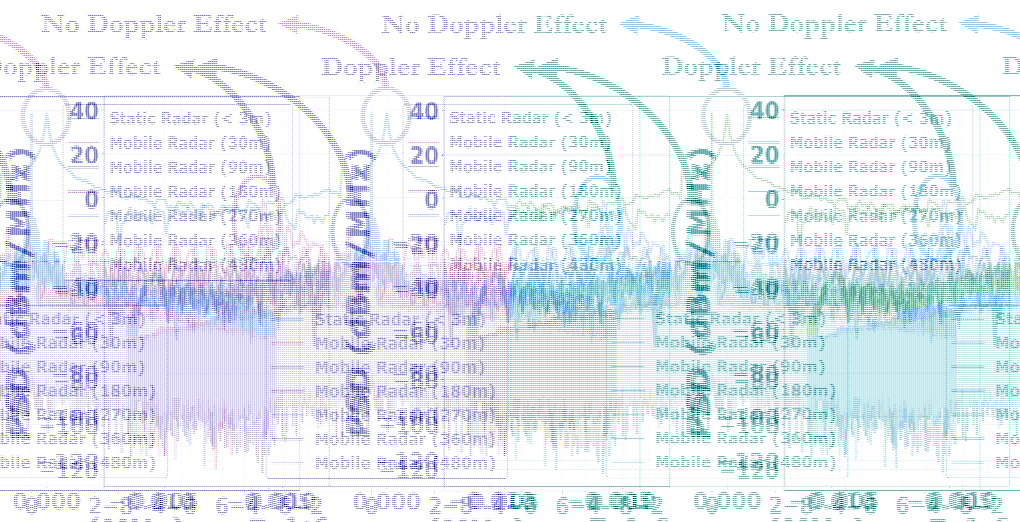}
    \caption{(a) Comparison of PSDs between static and mobile radars. (b) Doppler-induced spectral broadening observed at higher FFT resolutions.}
    \label{fig:validate_doppler}
    \vspace{-0.1in}
\end{figure}

\vspace{-0.05in}
Figure~\ref{fig:validate_doppler} demonstrates the impact of Doppler induced by airborne mobility. Spectral broadening is visible in PSD plots (Figure~\ref{fig:validate_doppler}(a)) and becomes more pronounced with increased FFT resolution (Figure~\ref{fig:validate_doppler}(b)). Figure~\ref{fig:airborne_eval}(b) further confirms that higher speeds induce sharper Doppler spread, validating the fidelity of our emulation setup.

\begin{table}[!b]
\vspace{-0.12in}
\centering
\caption{Dataset Collection Using PROPSIM Under Varying Channel and Mobility Conditions}
\label{tab:dataset-collection}
\renewcommand{\arraystretch}{1.1}
\scriptsize
\begin{tabular}{|p{2.3cm}|p{4.2cm}|p{1.3cm}|}
\hline
\textbf{Channel Model} & \textbf{Scenario} & \textbf{RADAR Speed (m/s)} \\
\hline
High Speed Train - Line Of Sight & Radar @ [30, 90, 180, 270, 360, 480] m from Base Station, OFDM fixed @ 180 m & 200 \\
\hline
Urban Microcell (UMi) - Line Of Sight & UEs and radar placed at same horizontal distances, OFDM fixed @ 180 m & 120 \\
\hline
Urban Microcell (UMi) - Non-Line Of Sight & NLOS propagation with fixed UE and radar layout, OFDM fixed @ 180 m & 120 \\
\hline
UMi - Clustered Delay Line - C & Radar at 6 altitudes (2–100 m), varied distances, OFDM fixed @ 180 m & 100 \\
\hline
Urban Macrocell - Line Of Sight & Radar at fixed distance 30 m, speeds varied: 25–150 m/s, OFDM fixed @ 180 m & 25–150 \\
\hline
\end{tabular}
\end{table}
\vspace{-0.05in}
\subsection{Airborne Radar Detection/Localization Performance}
\begin{figure}[!h]
    \centering
    \vspace{-0.1in}
    \includegraphics[width=0.49\textwidth]{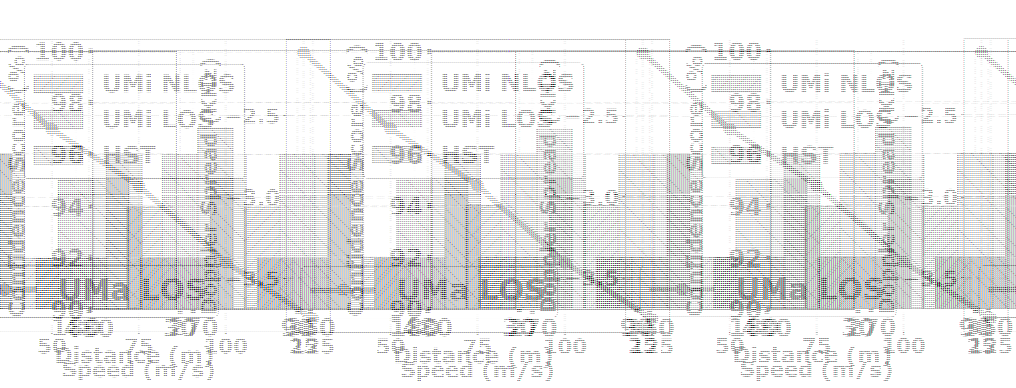}
    \vspace{-0.2in}
    \caption{(a) Confidence scores across distance and channel conditions. (b) Doppler spread trend increases with radar speed.}
    \label{fig:airborne_eval}
    \vspace{-0.15in}
\end{figure}

\vspace{-0.07in}
We retrain the spectrogram-based radar localization model introduced in Section~\ref{radar_localization} using the augmented dataset generated from the airborne testbed. Spectrograms are generated from the I/Q samples collected across different channel models and mobility profiles. Figure~\ref{fig:airborne_eval}(a) shows the model’s confidence probabilities or confidence scores in detecting airborne radar across varying distances and channel conditions, maintaining high detection probabilities across most 
scenarios.

\vspace{-0.05in}
\subsection{Discussion}
\vspace{-0.05in}
Figures~\ref{fig:airborne_testbed}–\ref{fig:airborne_eval} collectively illustrate the feasibility of the O-DSS framework in detecting airborne radars under dynamic mobility conditions. The PROPSIM-enabled testbed captures realistic Doppler and fading behaviors, validated through PSD and FFT-based analyses. The retrained model maintains high confidence level, demonstrating the framework’s adaptability to high-mobility environments. 

While our initial evaluations confirm the feasibility of airborne radar detection using O-DSS, the future work will involve comprehensive studies on end-to-end coexistence strategies under varying trajectories, altitudes, and multi-radar interference conditions to fully characterize and optimize O-DSS performance in both shipborne and airborne scenarios.
\vspace{-0.08in}

\vspace{-0.02in}
\section{Conclusions}
This paper presents O-DSS, an O-RAN compliant dynamic spectrum sharing framework for real-time cellular-radar coexistence in mid-band frequencies. By combining KPM based radar detection with spectrogram based localization, O-DSS achieves fast, low-overhead incumbent protection while enabling intelligent spectrum control through PRB blanking and radar-aware MCS adaptation. A key contribution is our open dataset covering both terrestrial and airborne radar scenarios. Evaluations simulations and OTA trials show O-DSS achieves sub-60~ms detection and  $\sim$700~ms evacuation, outperforming prior works and regulatory baselines. Crucially, O-DSS maintains low BLER and maintains throughput, achieving not just incumbent protection, but true spectrum coexistence.

The authors have provided public access to their data and code at \url{https://www.nextgwirelesslab.org/datasets}

\bibliographystyle{IEEEtran}
\bibliography{reference}

\end{document}